\documentclass[conference]{IEEEtran}
\pdfoutput=1

\IEEEoverridecommandlockouts
\usepackage{cite}
\usepackage{amsmath,amssymb,amsfonts}
\usepackage{algorithmic}
\usepackage{graphicx}
\usepackage{textcomp}
\usepackage{xcolor}
\def\BibTeX{{\rm B\kern-.05em{\sc i\kern-.025em b}\kern-.08em
    T\kern-.1667em\lower.7ex\hbox{E}\kern-.125emX}}

\makeatletter
\newcommand{\linebreakand}{%
  \end{@IEEEauthorhalign}
  \hfill\mbox{}\par
  \mbox{}\hfill\begin{@IEEEauthorhalign}
}
\makeatother

\title{Mitigating Backdoor Attacks in Federated Learning}

\author{\IEEEauthorblockN{Chen Wu}
\IEEEauthorblockA{\textit{Computer Science and Engineering} \\
\textit{Pennsylvania State University}\\
cvw5218@psu.edu}
\and
\IEEEauthorblockN{Xian Yang}
\IEEEauthorblockA{\textit{Computer Science} \\
\textit{North Carolina State University}\\
xyang45@ncsu.edu}
\linebreakand 
\IEEEauthorblockN{Sencun Zhu}
\IEEEauthorblockA{\textit{Computer Science and Engineering} \\
\textit{Pennsylvania State University}\\
sxz16@psu.edu}
\and
\IEEEauthorblockN{Prasenjit Mitra}
\IEEEauthorblockA{\textit{Information Sciences and Technology} \\
\textit{Pennsylvania State University}\\
pum10@psu.edu}
}

\begin{document}

\maketitle

\begin{abstract}
Malicious clients can attack federated learning systems using malicious data, including backdoor samples, during the training phase.
The compromised global model will perform well on the validation dataset designed for the task, but a small subset of data with backdoor patterns may trigger the model to make a wrong prediction.
There has been an arms race between attackers who tried to conceal attacks and defenders who tried to detect attacks during the aggregation stage of training on the server-side.
In this work, we propose a new and effective method to mitigate backdoor attacks \textit{after} the training phase.
Specifically, we design a federated pruning method to remove redundant neurons in the network and then adjust the model's extreme weight values. 
Our experiments conducted on distributed Fashion-MNIST show that our method can reduce the average attack success rate from 99.7\% to 1.9\% with a 5.5\% loss of test accuracy on the validation dataset.
To minimize the pruning influence on test accuracy, we can fine-tune after pruning, and the attack success rate drops to 6.4\%, with only a 1.7\% loss of test accuracy.
Further experiments under Distributed Backdoor Attacks on CIFAR-10 also show promising results that the average attack success rate drops more than 70\% with less than 2\% loss of test accuracy on the validation dataset. 
\end{abstract}

\begin{IEEEkeywords}
federated learning, backdoor attack, federated model pruning, machine-learning security
\end{IEEEkeywords}

\section{Introduction}
The success of deep learning models relies on large-scale datasets.
Traditional training methods require collecting the training data and centralizing the data in one machine or a data center. Nowadays, people are getting more sensitive and careful about sharing data with others. It becomes much harder and more expensive to gather data from different sources compared with the old days. 
To solve this problem, researchers proposed a new training method called federated learning \cite{DBLP:conf/aistats/McMahanMRHA17} to enable collaborative model training without sharing datasets among different clients or with the server. 

However, federated learning systems~\cite{DBLP:conf/aistats/McMahanMRHA17} are vulnerable to attacks from malicious clients \cite{DBLP:conf/icml/BhagojiCMC19, DBLP:journals/corr/abs-1807-00459, DBLP:conf/nips/BaruchBG19, DBLP:conf/uss/FangCJG20, mhamdi2018hidden}. 
The server does not have access to the data of clients, so it cannot verify model updates from clients, especially when the system is augmented with secure aggregation protocols to further protect the client privacy \cite{DBLP:conf/ccs/BonawitzIKMMPRS17}.
Theoretically, malicious clients can send any updates to the server and the server's global model could be easily compromised if the server has no effective protection to identify/process those malicious updates.

Existing defense methods focus on the federated aggregation process where the server receives model updates from all the clients. These methods try to distinguish malicious updates from benign ones. Byzantine-robust aggregation rules, such as Krum \cite{DBLP:conf/nips/BlanchardMGS17}, Bulyan \cite{DBLP:conf/icml/MhamdiGR18}, trimmed mean \cite{DBLP:conf/icml/YinCRB18} and median \cite{DBLP:conf/icml/YinCRB18}, all use statistical characteristics of model weights.
However, they have failed to detect backdoor attacks in federated learning \cite{DBLP:journals/corr/abs-1807-00459, DBLP:conf/icml/BhagojiCMC19, DBLP:conf/iclr/XieHCL20, DBLP:journals/corr/abs-1808-04866, DBLP:conf/uai/XieKG19} because the non-IID distribution of data among different clients creates enough space for the attacker to hide malicious updates from being detected. 

Backdoor attacks often trigger ``backdoor neurons'' which are the neurons that are activated only in the presence of backdoor-ed images  \cite{DBLP:journals/corr/abs-1708-06733}. 
Studies \cite{DBLP:conf/raid/0017DG18, DBLP:conf/sp/WangYSLVZZ19} have shown that pruning those ``backdoor neurons'' could greatly mitigate backdoor attacks without hurting too much model performance. However, these pruning methods cannot be directly used in our case because they rely on reliable sources of ``clean'' data, which is not guaranteed in federated learning scenarios (which is designed to protect the privacy of clients' data). 
To address this problem, we propose a novel federated pruning process that does not require access to the clients' original data and helps in deciding the pruning sequence of neurons. 

We introduce two federated pruning methods by eliminating neurons that are not being activated by clients' inputs. Our federated pruning method requires clients to rank the dormant level of neurons in the network. 
Basically, neurons that are not frequently activated by the network are called dormant neurons.
After aggregation, the server will determine the dormant neurons based on the information provided by all the clients and steadily prune the neural networks while checking the model performance on a small validation dataset. 

Our empirical studies have shown that a single federated pruning process is insufficient to thoroughly eliminate backdoor attacks. The success of the federated pruning method highly depends on the attacker's targets. For example, the pruning process can fully eliminate backdoor behavior when the attacker tries to backdoor digit number 9 to number 1, while the same pruning process fails
when the attacker tries to backdoor 9 to 5. The reason is that in some cases, benign and backdoor behaviors utilize the same set of neurons. Thus, pruning those neurons will degrade the performance of the model on both clean data and backdoored instances. 

Despite the challenges, we observe that 
under the majority assumption
when the number of neurons/features supporting correct labels is more than the number of neurons/features supporting backdoor labels, a malicious client has to 
introduce extreme values for the inputs or extreme values for the weights of the neurons to reverse the correct prediction results to backdoor labels . 
Thus, by limiting the inputs and the weights of the neurons, we can mitigate the backdoor attacks. 

The contributions of this paper are as follows:
\begin{itemize} \setlength\itemsep{0em}
    \item We propose two federated pruning methods to remove redundant neurons in the network without degrading accuracy significantly in a scenario where we do not have access to the clients' private dataset.
    
    \item Our experiments show that the defensive effect of pruning neurons highly depends on the target label attacked by the malicious clients. Thus, we propose to adjust extreme weights in the network to degrade the backdoor attacks. Experiments on MNIST and Fashion-MNIST have shown that our method can effectively reduce the average attack success rate from over $99\%$ to be less than $2\%$. Experiments on CIFAR-10 under state-of-the-art distributed backdoor attacks have also shown that our method can reduce the average attack success rate more than $70\%$ with a loss of test accuracy less than $2\%$.
    
    \item We propose a federated fine-tuning process after neuron pruning to improve the model performance on the validation dataset. Experiments show that adjusting extreme weights after fine-tuning can degrade the attack success rate to less than $10\%$ while the global model's accuracy rate on the validation dataset can improve $5\%$. 
    
\end{itemize}
 
\section{Related Works}
\subsection{Backdoor Federated Learning}
Bagdasaryan et al. \cite{DBLP:journals/corr/abs-1807-00459} introduced the ``semantic backdoor'' concept that uses rare features in the real world as a trigger and does not require the attacker to modify the input of the model at inference time. For example, an adversary could compromise the global model by predicting car images with racing stripes as birds, whereas other car images would still be predicted as cars. Sun et al. \cite{DBLP:journals/corr/abs-1911-07963} further showed that allowing the non-malicious clients to have correctly labeled samples from the targeted tasks could not prevent such backdoor attacks. Bhagoji et al. \cite{DBLP:conf/icml/BhagojiCMC19} showed that with 10\% of the clients being compromised, a backdoor can be introduced by poisoning the model sent back to the server, even with the presence of anomaly detectors or Byzantine-resilient aggregation mechanisms used by the server. 

\subsection{Attacks and Defenses in Federated Learning}
Previously proposed byzantine-robust aggregation rules, like Krum \cite{DBLP:conf/nips/BlanchardMGS17}, Bulyan \cite{DBLP:conf/icml/MhamdiGR18}, trimmed mean \cite{DBLP:conf/icml/YinCRB18} and median \cite{DBLP:conf/icml/YinCRB18}, all using statistical characteristics of model weights, were reported to have failed to detect backdoor attacks in federated learning \cite{DBLP:journals/corr/abs-1807-00459, DBLP:conf/icml/BhagojiCMC19}. Non-IID distribution of data gives attackers enough space to forge their backdoor-model updates. Fang et al. \cite{DBLP:journals/corr/abs-1911-11815} have further shown an \textit{untargeted attack} that by directly manipulating the local model parameters on the compromised client devices during the learning process, the global models with previous byzantine-robust aggregation rules would suffer a worse testing error rate. To better conceal attackers' updates from being detected, Xie et al. \cite{DBLP:conf/iclr/XieHCL20} proposed a distributed backdoor attack method that decomposes a global trigger pattern into separate local patterns and embeds them into the training set of different attackers. 

Li et al. \cite{DBLP:journals/corr/abs-2002-00211} proposed a spectral anomaly detection based framework that detects the abnormal model updates based on their low-dimensional embeddings, in which the noisy and irrelevant features are removed while the essential features are retained. They showed that in low-dimensional latent feature space, the abnormal (malicious) model updates can be easily differentiated from the normal updates. 
Most existing defense methods are trying to distinguish attackers' updates from benign clients' updates, while the attackers are trying to modify their updates as close to the other updates as possible. It is hard to tell which method is more effective under different datasets and various data distributions. 
Our work differs from the existing ones by focusing on  mitigating backdoor attacks further after the training phase.

\subsection{Pruning Against Backdoor Attacks}
Gu et al. \cite{DBLP:journals/corr/abs-1708-06733} showed that poisoned data designed to introduce a backdoor often triggers ``backdoor neurons". Based on this assumption, pruning defenses \cite{DBLP:conf/raid/0017DG18, DBLP:conf/sp/WangYSLVZZ19} attempt to remove activation units that are inactive on clean data. This method requires ``clean'' data that is representative of the global dataset being used to identify those ``backdoor neurons'', and such ``clean'' data is typically not approachable by the server in federated learning scenarios. 
Different from the above work, our work proposes a federated neuron pruning method. 

\begin{figure*}
    \centering
    \includegraphics[width=1\textwidth]{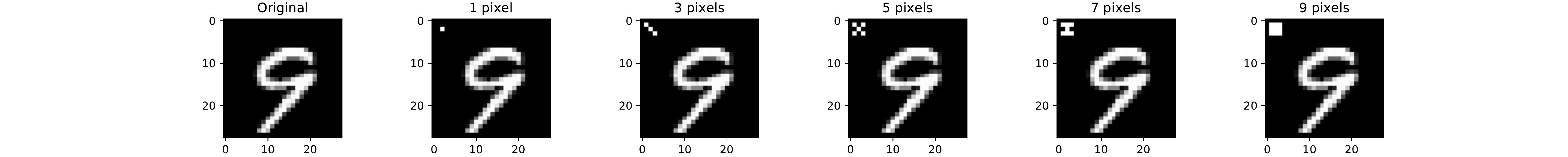}
    \vspace{-4ex}
    \caption{\scriptsize An original image from the MNIST dataset, and the backdoored version of this image using different pixels backdoor pattern.}
    \label{fig:attack_patterns_graph}
    \vspace{-2ex}
\end{figure*}

\section{Problem Definition}
In this section, we introduce our federated learning environment settings, including the aggregation rule, data distribution among clients, and backdoor attack methods employed in this paper.

\subsection{Global Model Learning}
According to the federated learning definition proposed by McMahan et al. \cite{DBLP:conf/aistats/McMahanMRHA17}, training data is from a number of clients and cannot be shared with others other than the client himself.
We assume the total number of clients is $N$ and each client has $n_{i}$ number of samples. At each time $t$, the server randomly selects a portion of clients $k \cdot N$ ($0 < k \leq 1$) from all clients and asks them to train the global model on their local dataset. Then, the \textit{FedAvg} algorithm \cite{DBLP:conf/aistats/McMahanMRHA17} will update the global model by aggregating the weight updates from these clients as shown below.
\vspace{-0.5ex}
\[
    \omega_{t+1} = \omega_{t} + \eta \cdot \frac{\sum_{i = 1}^{kN} n_{i} \Delta\omega_{t+1}^{i}}{\sum_{i=1}^{kN} n_{i}}
    \vspace{-1ex}
\]

where $\omega_{t}$ is the previous model parameters at time $t$, $\omega_{t+1}$ is the updated model parameters at time $t+1$, $\Delta\omega_{t+1}^{i}$ is the updated changes of model parameters provided by client $i$, $\eta$ is the global model learning rate. 
In this paper, since our defense method is not focused on the aggregation process, we simplify the above rules to speed up the backdoor attack and make sure the attack is successful. 
To be more specific, we make some assumptions to simplify the above learning process. Firstly, we set the number of samples $n_{i}$ same for every client, because otherwise the attacker can simply strengthen their updates by claiming that they have a larger number of samples to overwhelm the updates from others. Secondly, to improve the global training speed and better monitor the attacking process behavior, we set the same learning rate $\eta_{i}$ for each client (including the attacker). Thirdly, we assume all clients will participate in every round of aggregation so that the random selection of clients process will not affect the performance of the global model and attacks. The global model aggregation rule is simplified as follows.
\vspace{-2ex}
\[
    \omega_{t+1} = \omega_{t} + \frac{1}{N} \sum_{i=1}^{N} \Delta\omega_{t+1}^{i}
    \vspace{-1ex}
\]

\subsection{Threat Model}
The goal of the attacker is to ensure the final global model only fails on a specific task while performing reasonably well on the other tasks.
For example, on the MNIST dataset \cite{lecun1998gradient}, the attacker would like to make the global model mispredict images with backdoor patterns as another label selected by the attacker, while the other normal images would still be predicted correctly. Consequently, the server can hardly recognize that the global model has been compromised even with a substantial evaluation dataset. 
In general, the attacker would like the global model to predict all testing samples $\{x_{i}\}$ with correct labels $y_{i} = T$ correctly, while corrupted samples with backdoor patterns $\{x_{i}'\}$ will be predicted to a wrong label $F$, where label $T$ and label $F$ can both be selected by the attacker.

We make the following assumptions about the attacker in our experiments: (i) According to our federated learning scenario, all the clients participate in every round of model aggregation. So, there will be at least one attacker in each training iteration. (ii) The malicious client only has access to his own dataset and has no access to the testing dataset. The backdoor training samples are created by applying backdoor patterns to the local dataset owned by the attacker.

\subsection{Backdoor Attack}

In this paper, we start with similar backdoor patterns as in BadNets \cite{DBLP:journals/corr/abs-1708-06733}, which changes some pixels in a picture to form a pattern. During the local training phase, the attacker would train the model with both original images and the backdoored version of those images at the same time. In this way, he can enforce the model to learn the backdoored pattern instead of misclassifying both the original images and the backdoored images. For example, among the images in Fig \ref{fig:attack_patterns_graph}, the original image would still be predicted as `9', while the backdoored images are predicted as `1' or any other label chosen by the attacker.

\subsection{Model Replacement Attack}
After local training, we use the \textit{model replacement attack} \cite{DBLP:journals/corr/abs-1807-00459} to make sure the updates from the attacker will survive the \textit{FedAvg} aggregation on the server's side.
The basic idea behind \textit{model replacement attack} is that the attacker wants the global model to be as close to his local trained model as possible. In the ideal situation, the global model will be completely replaced by the attacker's model as shown below:
\vspace{-1ex}
\begin{equation}
\label{eq:model_replace_atk}
    x_{atk} = \omega_{t+1} = \omega_{t} + \frac{1}{N} \sum_{i=1}^{N} (x_{t+1}^{i} - \omega_{t})
    \vspace{-1ex}
\end{equation}

where $x_{atk}$ is the attacker's model, $\omega_{t+1}$ is the global model at time $t+1$, $x_{t+1}^{i}$ is the local model trained by client $i$ at time $t+1$, $N$ is the number of all clients that participate in the federated learning process. So this function means that the global model $\omega$ at time $t+1$ is an averaged mean of model updates from $N$ clients at time $t+1$ and the goal of the attacker is to replace this global model $\omega$ at time $t+1$ with attacker's model $x_{atk}$. 
Let $x^{m}_{t+1}$ be the update from the malicious client $m$ at time $t+1$. Then, by solving Equation \ref{eq:model_replace_atk}, we can have the following:
\vspace{-2ex}
\[
\label{eq:model_replace_atk_update}
    x_{t+1}^{m} = N \cdot x_{atk} - N \cdot \omega_{t} - \sum_{i=1}^{N-1} (x_{t+1}^{i} - \omega_{t}) + \omega_{t}
    \vspace{-1ex}
\]

As assumed in the work of Bagdasaryan et al. \cite{DBLP:journals/corr/abs-1807-00459}, $\sum_{i=1}^{N-1} (x_{t+1}^{i} - \omega_{t}) \approx 0$. With the convergence of global model, these deviations cancel out, so the attacker's update can be simplified as following:
\vspace{-1ex}
\[
\label{eq:model_replace_simple_update}
    x_{t+1}^{m} = N \cdot (x_{atk} - \omega_{t}) + \omega_{t}
    \vspace{-1ex}
\]

\begin{figure*}
    \centering
    \includegraphics[width=1.0\textwidth]{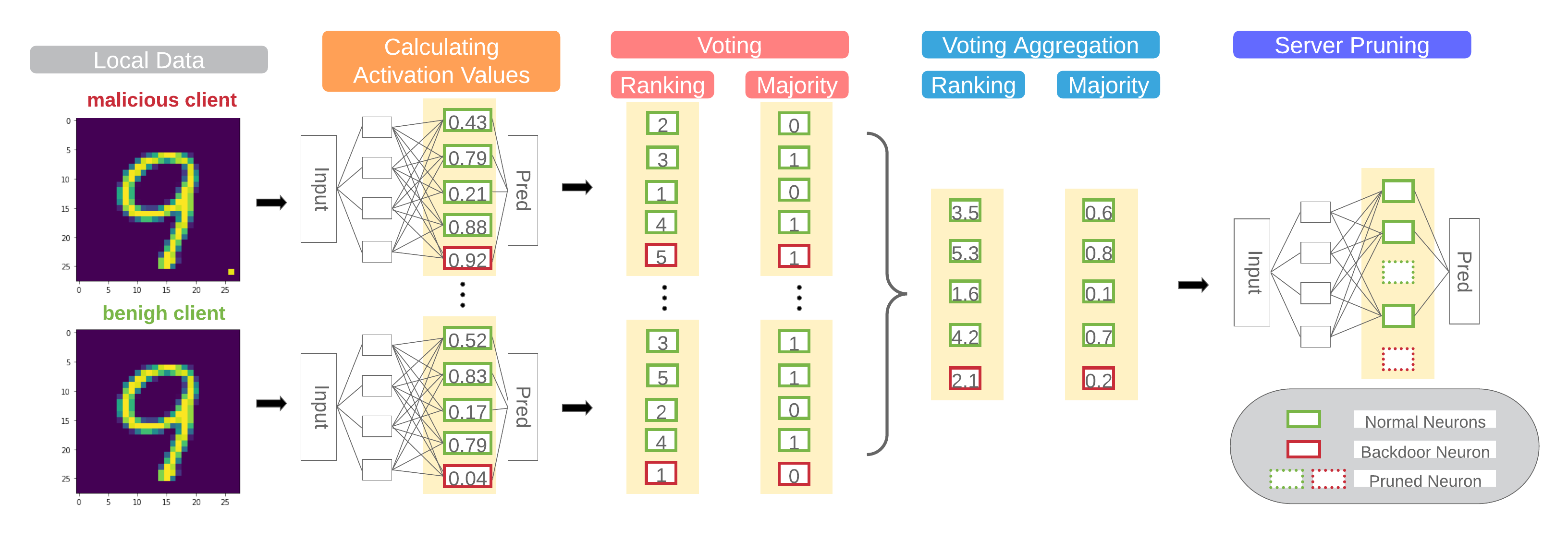}
    \vspace{-6ex}
    \caption{Illustration of the federated pruning process.}
    \label{fig:federated_pruning_illustration}
    \vspace{-2ex}
\end{figure*}

However, since the deviations do not cancel out in the training process according to our experiments, we replace $N$ in the above equation with an attack update amplification coefficient $\alpha$ ($1 \leq \alpha \leq N$). We find that the value $\alpha$ is related to the data distribution among clients. If the distribution of dataset is more similar among clients, a higher value of $\alpha$ is needed to achieve better results on the backdoor tasks. 


\section{Defense Method}
Empirical studies by Gu et al. \cite{DBLP:journals/corr/abs-1708-06733} showed that the backdoor inputs will trigger the neurons in a DNN that are typically not used by normal clean inputs. These so-called ``backdoor neurons'' are leveraged by the attacker to recognize backdoor patterns and trigger misbehavior while keeping silent when the input data is clean. 
Based on this finding, Liu et al. \cite{DBLP:conf/raid/0017DG18} proposed a Fine-Pruning method to remove the ``backdoor neurons'' in the DNN, thus mitigating backdoor attacks. 
The key idea behind this pruning method is to find the dormant neurons that are not frequently activated in the network.
By using clean training inputs to activate the DNN, they record the averaged activation values of each neuron in the last convolutional layer of the network. 
Then, they can iteratively prune neurons based on the increasing order of averaged activation values and stop before the accuracy on the validation dataset drops below a certain threshold.



However, in federated learning, the server does not have access to the training data of clients. Also, there is no way to guarantee a clean source of training inputs to activate the network, since attackers hide among the clients.
So, in this paper, we propose a distributed way of pruning neurons, as shown in Fig \ref{fig:federated_pruning_illustration}. 
The first step is to ask all the clients to record the averaged activation value of each neuron based on their local training dataset. 
Then, each client will determine a local pruning sequence based on the collected information and send this local pruning sequence to the server.
The server will gather such information from all clients to further determine a global pruning sequence indicating which neurons are less used and thus can be pruned first. 
It can test on a small validation dataset to decide the pruning rate (how many neurons need to be pruned) following this pruning sequence and select the maximum pruning rate with an acceptable test accuracy on the validation dataset.
On the other hand, if the server does not have such a validation dataset or its dataset is too small and not able to represent the data distribution in the real world, 
it can pass the global pruning sequence back to each client and ask them to report the test accuracy on each client's dataset under different pruning rates. The server will collect this feedback and determine the final pruning strategy. 

There are two advantages to this method. Firstly, this method does not require access to the clients' datasets and thus protects the privacy of each client. 
Secondly, it is computationally efficient in federated learning since it only needs one round of communication between server and clients if the server has a validation dataset. Otherwise, an extra round of communication is required to collect the clients' feedback on the pruning sequence. 
It is even possible to include more clients to participate in this step than in the training process. 
Along with the advantages, it also comes with shortcomings. For example, we can no longer have guaranteed ``clean data''. Since the attackers hide among the clients, they can return manipulated updates to maintain the attack success rate. Thus, we further propose two algorithms to minimize the influence from a minority group of attackers.

\subsection{First Approach: Ranking Vote}
The ideal situation in federated pruning is that the server can directly collect all the average activation records from each client. However, directly passing the real values may result in both privacy and security problems. For privacy concerns, real values may reveal some information about the clients' original dataset. For security concerns, the attackers may easily manipulate the final aggregation results by changing the updated values just as they did in the model learning process. 

To solve this problem, we propose a ranking vote algorithm, where each client provides a ranking of all the neurons in the last convolutional layer based on their averaged activation values. 
We consider each neuron stands for a sliding window in the convolutional layer. 
The sum of the output from that sliding window will be treated as the neuron's activation value. 
If a convolutional layer has an output size of $(B, C, H, W)$, where $B$ is the batch size, $C$  number of channels (sliding windows), $H$ height of input planes in pixels, and $W$ width in pixels, then we say this convolutional layer has $C$ neurons and the activation value of each neuron is the sum of $(H, W)$ outputs for each channel in $C$. 
The averaged activation value of each neuron will be the sum of all the activation values divided by the number of images used in the training. 
Each client will rank these neurons based on their averaged activation values in ascending order from 1 to $C$. 
In the next step, the server can create a global dormant ranking of all the neurons by averaging the ranking positions of individual neurons from all the clients. 
If we denote the ranking position of neuron $k$ provided by client $i$ as $R_{k}^{i}$, the aggregation of ranking for each neuron $k$ can be represented as $R_{k} = \frac{1}{N} \sum_{i=1}^{N} R_{k}^{i}$. 
In the end, the server iteratively prunes neurons in the decreasing order of the global dormant ranking until the accuracy on the validation dataset drops below a certain threshold. 

On the other hand, if the server does not have sufficient representative validation dataset to determine when to stop pruning, it can send the global neuron ranking back to each client and ask them to test on clients' local datasets following the same pruning sequence. After collecting the pruning results of each client, the server can decide how many neurons are going to be pruned following this sequence. 

\subsection{Second Approach: Majority Vote}
\label{sec:fed_pruning_baseline_approach}

In the above Ranking Vote algorithm, with more neurons in the model, malicious updates of ranking information will have larger influence on the server's global ranking. To further limit the impact of potential attackers, we can simplify the updates from the clients to be only zero or one for each neuron. 
The value \textit{one} indicates that the neuron needs to be kept while  \textit{zero} means the neuron needs to be pruned from the view of this client.
Then, it would be much harder for the attackers to manipulate the final results if benign users are the majority group. Also, this method can better protect the clients' privacy by revealing less information about the activation records on the local dataset.

Specifically, the server will provide a pruning rate , say $p\%$, to all the clients. Each client needs to decide which neurons should be pruned based on their averaged activation records and the pruning rate. If the average activation value of a neuron belongs to the smallest value group (i.e., the value is below $p\%$ of all the values), it will be pruned. If the neuron should be pruned, it will be assigned value `0'; otherwise, it will be assigned `1'. At this point, we can imagine that each client will create a mask for all neurons and the mask only has values `0' and `1'. Then, the server will aggregate all the masks from all clients to have a majority vote information about each neuron. Neurons with higher values stand for their importance among clients; neurons with lower values stand for their dormancy among clients. In the last step, the server can prune neurons in increasing order of their aggregated values until the accuracy on the validation dataset drops below a certain threshold. 

With the improvement of security and privacy, this method may require more rounds of communication between clients and the server compared with the Ranking Vote algorithm. Because the pruning rate $p\%$ can hardly be selected without any prior knowledge. Although our experiments show that the pruning rate between $30\%$ and $70\%$ performs well in various situations, the server may still need to try different pruning rates in a real situation. Thus, extra rounds of calculation and communication between clients and the server may be needed.

\subsection{Adjusting Extreme Weights}
In our empirical study, we find that the pruned network alone cannot guarantee the mitigation of backdoor attacks. The effect of pruning is highly dependent on the data distribution among clients and the target labels chosen by the attackers. 
Despite the challenges, we observe that under the majority assumption when the number of neurons/features supporting correct labels is more than the number of neurons/features supporting backdoor labels, a malicious client has to introduce extreme values for the inputs or extreme values for the weights of the neurons to reverse the correct prediction results to backdoor labels. Thus, by limiting the inputs and the weights of the neurons, we can mitigate the backdoor attacks.

To limit the input ranges, we make an input normalization step for all the inputs to the model. 
To adjust extreme weight values in the network, we scan all the weights in the last convolutional layer and zero the weights that are larger or smaller than thresholds based on the mean and standard deviation of the weights in that layer. 
For example, if the mean value of all the weights in layer $i$ is $\mu_{i}$ and the standard deviation is $\sigma_{i}$. Then, we set the threshold $s = \mu_{i} \pm \Delta \cdot \sigma_{i}$ that controls the valid range of all the weights in this layer. 
$\Delta$ is a server-defined hyperparameter in this function.
We will discuss more about the selection of $\Delta$ values in the Experiments section.
Finally, all the weights that are beyond the threshold $s = \mu_{i} \pm \Delta \cdot \sigma_{i}$ will be set to zero. 
Our empirical studies have shown that this process is effective in reducing the backdoor task success rate to a low level after model pruning.

\subsection{Fine-tuning}
Pruning neurons and adjusting extreme weights will cause a drop in the model's test accuracy on the validation dataset. We neither want our model to be vulnerable under backdoor attacks nor expect the model to perform poorly on designed tasks. So, we propose to employ the following fine-tuning process after pruning neurons. 
Specifically, the server will send the pruned neural network back to clients and ask for training updates. 
Each client will train the pruned network using its local data again. 
Just as we did in the training process, the server will update the global model by aggregating clients' weight updates. This process is called fine-tuning. Fine-tuning will continue for a few rounds until the pruned model recover the lost test accuracy (with the validation dataset) as much as possible, that is,
When the test accuracy does not improve any further or even starts to decrease, the server will stop the fine-tuning process.
In our experiments, it usually takes about ten rounds of updates. 
During this process, the attacker may also participate and try to push the backdoor attack success rate back.
To mitigate the attack, we perform an adjusting extreme weights process after fine-tuning. 
Empirical studies have shown that adjusting extreme weights can reduce the attack success rate back to a low level. 

\section{Experiments}
Our experiments are performed on MNIST \cite{lecun1998gradient}, Fashion-MNIST \cite{xiao2017/online} and CIFAR-10 \cite{krizhevsky2009learning} dataset with non-i.i.d. data distributions. We verified our defense method on backdoor tasks in different client data distribution, different backdoor tasks, different model architectures, and under different backdoor patterns. 
For experiments on MNIST, we use a model that consists of 2 convolutional layers and 2 fully connected layers. For experiments on Fashion-MNIST, we use a model that consists of 3 convolutional layers and 2 fully connected layers. For experiments on CIFAR-10, we use a model that consists of 4 convolutional layers and 3 fully connected layers. Further more, we applied the state-of-the-art \textit{Distributed Backdoor Attack} \cite{DBLP:conf/iclr/XieHCL20} on the CIFAR-10 experiments. 

\begin{figure}
    \vspace{-2ex}
    \centering
    \includegraphics[width=0.45\textwidth]{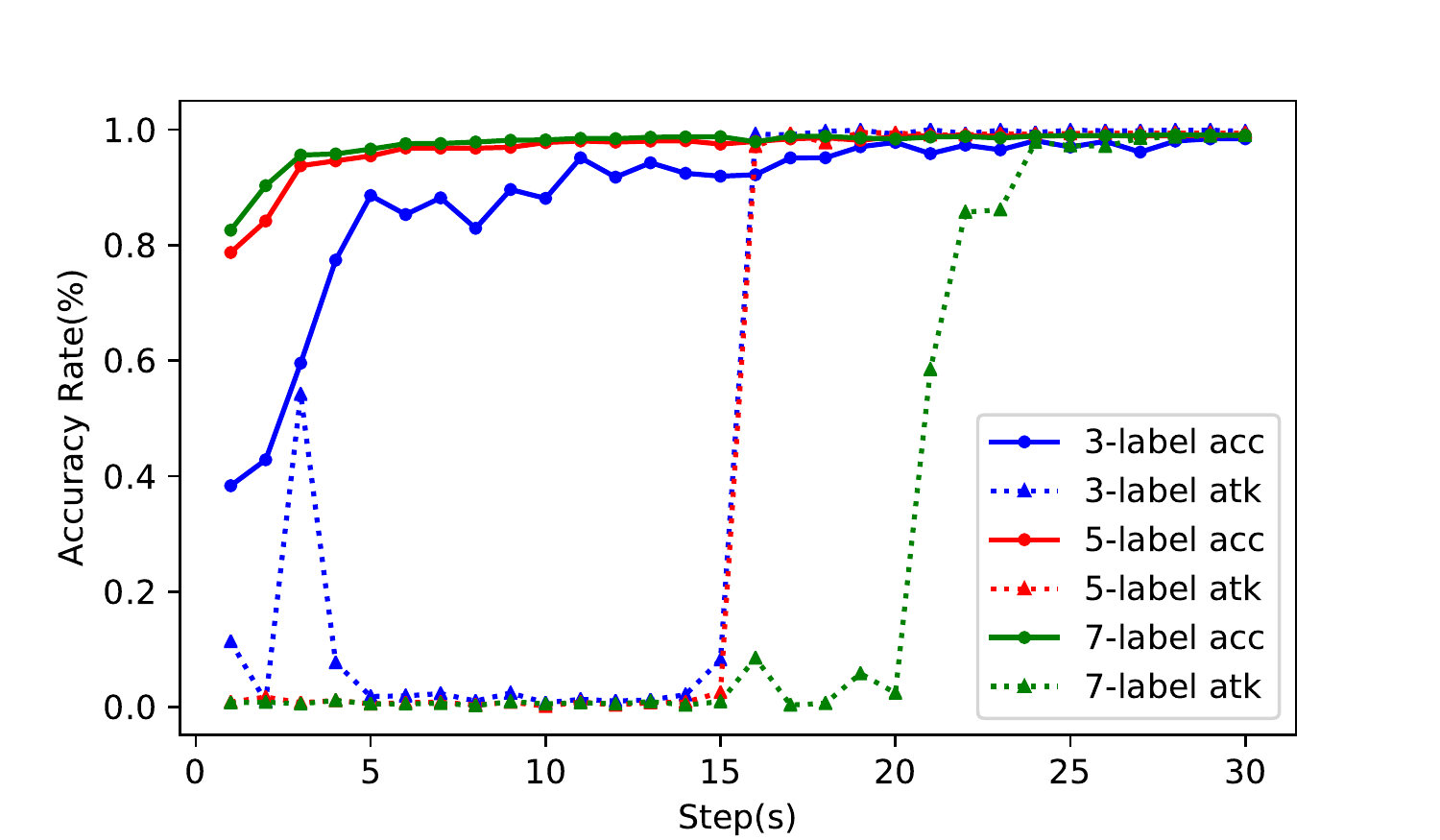}
    \vspace{-1ex}
    \caption{The training process of 3-label, 5-label and 7-label MNIST data distribution among 10 clients, tested on server's evaluation dataset. Solid line stands for the test accuracy on evaluation dataset, while dashed line stands for attack success rate on target labels.}
    \label{fig:data_distribution_train}
    \vspace{-2ex}
\end{figure}

\subsection{Client Data Distribution}
Our experiments show that the data distribution among clients can significantly influence the federated learning process. 
We define a $K$-label distribution among all the clients to simulate various real-world situations. $K$-label distribution means each client will be randomly assigned data belonging to $K$ different labels. 
For example, if $K = 10$, each client will have randomly assigned data from all the $10$ labels (MNIST dataset only has $10$ different labels, from digit ``$0$'' to digit ``$9$''). 
The data in each client follows the same uniform distribution, which means that every client will have roughly the same number of samples for each label. 
In another extreme case, if $K = 1$, each client will have data that belongs to a single label. Given the condition of $10$ clients, in this case each client will hold all the data belonging to a unique label. In this extreme situation, the training process would take much longer than the $10$-label distribution, and the backdoor attack would also be easier.
An example of the models' training performance under 3-label, 5-label, and 7-label distribution is in Figure \ref{fig:data_distribution_train}. 
It shows the change of test accuracy of the model on the evaluation dataset and the shift in backdoor attack success rate along with the number of rounds of training (X-axis).
Each point in the figure represents the performance of the global model at that round.
There's no defense method used in this process.
By the end of the training, both test accuracy on the evaluation dataset and backdoor attack success rate is higher than $98\%$. 
We can notice that the training of 3-label distribution is much more challenging (needs more rounds) than the 7-label distribution. In contrast, the attack to 3-label distribution is much easier (needs fewer rounds) than the 7-label distribution. A balanced distribution of data among clients is better for training and more robust under attacks.

\subsection{Neuron Pruning Methods Comparison}

\begin{figure}
    \vspace{-2ex}
    \centering
    \includegraphics[width=0.45\textwidth]{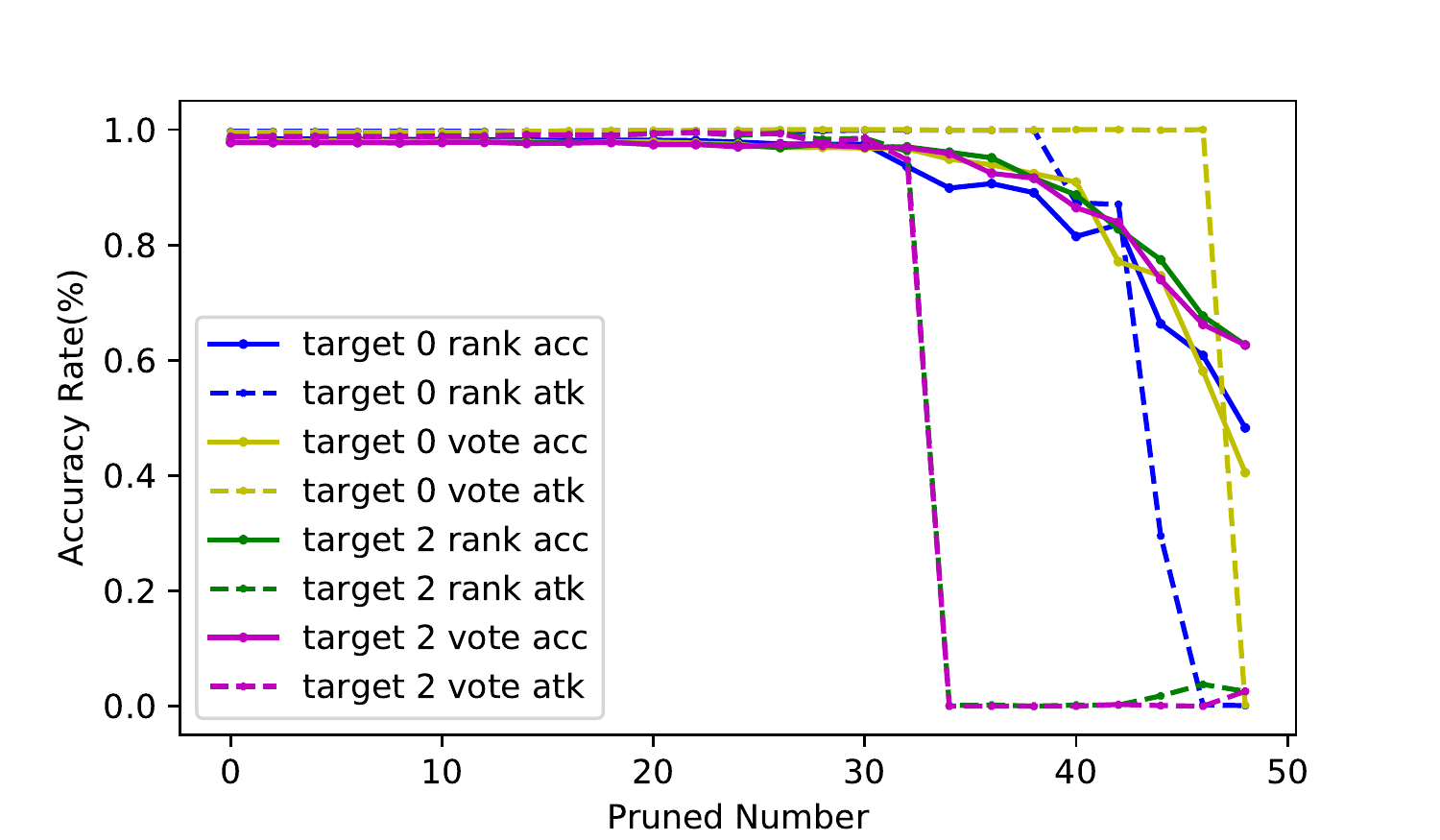}
    \vspace{-1ex}
    \caption{The neuron pruning process of models on MNIST dataset with 3-label distribution among 10 clients, tested on server's evaluation dataset. ``rank'' stands for using ranking vote algorithm and ``vote'' for using majority vote algorithm.
    Solid line stands for test accuracy on server's evaluation dataset, and dashed line stands for attack success rate on target labels.}
    \label{fig:neuron_pruning_methods_comp}
    \vspace{-2ex}
\end{figure}

In this paper, we proposed two algorithms to determine the pruning sequence of neurons: ranking vote and majority vote. Figure \ref{fig:neuron_pruning_methods_comp} shows their performance under different numbers of pruned neurons. There are 50 neurons in total. 
``target 0'' means the attacker wants to backdoor digit 9 as digit 0, and ``target 2'' means they want to backdoor digit 9 as digit 2.
As we can see that the pruning effect is almost identical between these two methods. 
More detailed experiment results will be reported in Table \ref{table:pruning_neurons_performance} in the Defense Method Evaluation section.

Another question that comes with this pruning process is when should we stop. As we can see in Figure \ref{fig:neuron_pruning_methods_comp}, when attacker's target label is 0, the attack success rate decreases after the test accuracy drops. Also, the server does not have backdoor images during testing. Hence, we can only decide the stop point based on the test performance. In this experiment, we determine the pruning stop point when the test accuracy drops over $1\%$ between two testing points. 


\begin{table*}[h]
\caption{Federated Pruning Performance under Different Attack Patterns}
\vspace{-2ex}
\centering
\begin{tabular}{|c|c|c|c|c|c|c|c|c|c|}
    \hline
    \multicolumn{1}{|l|}{Attack Pattern} & \multicolumn{2}{|l|}{Training Phase} & \multicolumn{3}{|l|}{Pruning Neurons} & \multicolumn{3}{|l|}{Adjusting Extreme Weights} \\ 
    \hline
    pixels & test acc & atk acc & num & test acc & atk acc & num & test acc & atk acc \\
    \hline
    1 & 98.5 & 99.9 & 22 & 97.3 & 98.7 & 131 & 97.2 & 0.4 \\
    3 & 98.5 & 100 & 30 & 96.6 & 100 & 139 & 97 & 34.8 \\ 
    5 & 98.4 & 100 & 34 & 97 & 100 & 138 & 95.2 & 1.4 \\ 
    7 & 98.5 & 100 & 30 & 96.2 & 96.9 & 138 & 96.8 & 32.9 \\ 
    9 & 98.4 & 99.8 & 30 & 97.6 & 2.3 & 133 & 96.2 & 0.5 \\ 
    \hline
\end{tabular}
\label{table:attack_patterns_performance}
\vspace{-2ex}
\end{table*}

\subsection{Extreme Value Threshold}
\begin{figure}
    \vspace{-2ex}
    \centering
    \includegraphics[width=0.45\textwidth]{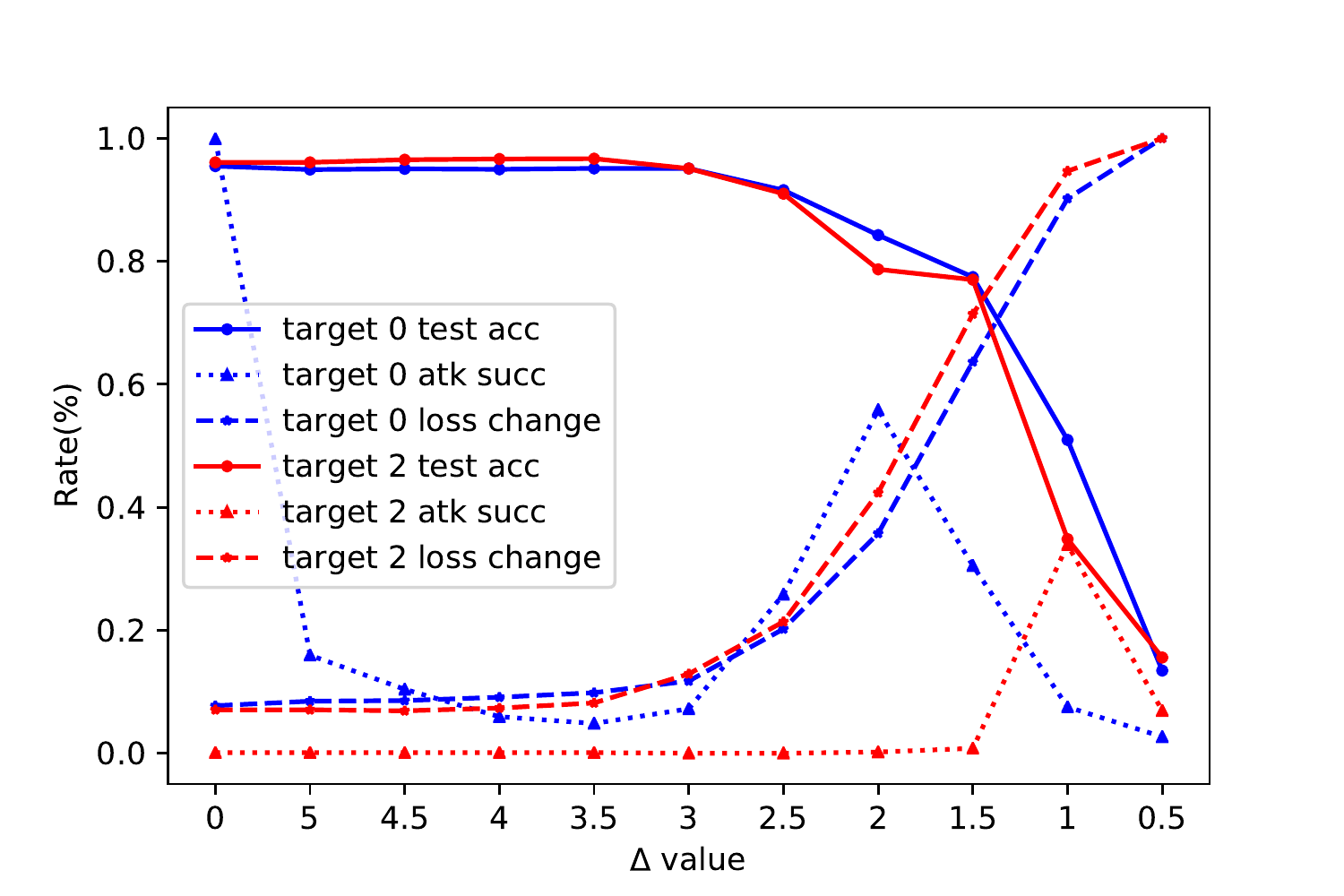}
    \vspace{-2.5ex}
    \caption{The extreme value pruning process of model with different pruning threshold defined by $\Delta$. The first data point in the graph, $\Delta=0$ stands for the original model without pruning extreme values. 
    The model is trained on MNIST dataset in 3-label distribution among 10 clients.}
    \label{fig:prune_extreme_values_plot}
    \vspace{-2ex}
\end{figure}

After pruning neurons in the network, the last step is to limit extreme weight values in the same layer that we prune neurons. We calculate the mean $\mu_{i}$ and the standard deviation $\sigma_{i}$ of all the weights in layer $i$. Then, we change all the weights that are beyond the threshold $s = \mu_{i} \pm \Delta \cdot \sigma_{i}$ to be zero. 
Figure \ref{fig:prune_extreme_values_plot} shows two examples that this pruning process can effectively mitigate backdoor attack without hurting the accuracy of the model on designed tasks. 
In one example, the attacker tries to backdoor digit 9 as digit 0. In the other example, the attacker tries to backdoor digit 9 as digit 2.
When the attack success rate is high at the beginning of this process, pruning extreme values can greatly decrease the attack success rate with even large $\Delta$ values. On the other hand, when the attack success rate is low (pruning neurons process already mitigate the backdoor attacks) at the beginning of this process, pruning extreme values will not increase the attack success rate. 
Typically, we can use the same stopping criteria as in the above pruning process. We can start the process with large $\Delta$ value and gradually decrease $\Delta$ value until the test accuracy drops below a certain threshold or the loss rises above a certain threshold.

\subsection{Backdoor Patterns}
We conduct experiments to study the impact of backdoor patterns on the attack and our defense method. 
As shown in Figure \ref{fig:attack_patterns_graph}, we implemented five different attack patterns with the backdoor task of changing the prediction results of digit 9 to digit 1 in the MNIST dataset. The experiment results of federated pruning performance under different attack patterns are shown in Table \ref{table:attack_patterns_performance}. 
An example of number of pixels attack can be seen in Figure \ref{fig:attack_patterns_graph}. \textit{test acc} stands for the performance of model on the test dataset. \textit{atk acc} stands for the performance of model on backdoor samples. \textit{num} in \textit{pruning neurons} stands for the number of neurons that are pruned during this process. \textit{num} in \textit{adjusting extreme weights} stands for the number of weights that are changed to zero in this process. The experiment data in this table is trained on a two layer convolutional neural network, with the first layer containing 20 neurons and the second layer containing 50 neurons. The backdoor task is trying to backdoor digit 9 in MNIST to digit 1.

In the extreme weights adjusting process, we fixed the threshold index $\Delta = 3$. 
Although we could achieve a much lower attack success rate in some patterns (3-pixels and 7-pixels attack patterns) when using methods, as shown in Figure \ref{fig:prune_extreme_values_plot}, by selecting the $\Delta$ value that test accuracy begins to decrease. To maintain a fair comparison between different attack patterns, we use the fixed $\Delta$ value to conduct the comparison experiment. 
We can observe that the structure of the attack pattern has some impact on the final attack performance. In other words, some patterns are more resistant to the neuron pruning process, and some are more resistant to the adjusting extreme weights process. 

\begin{table*}[h]
\caption{Neurons Pruning Process with Different Attack Targets}
\vspace{-2ex}
\centering
\begin{tabular}{|c|c|c|c|c|c|c|c|c|c|}
    \hline
    \multicolumn{2}{|l|}{Target} & \multicolumn{2}{|l|}{Training Phase} & \multicolumn{3}{|l|}{Pruning Neurons - Ranking Vote} & \multicolumn{3}{|l|}{Pruning Neurons - Majority Vote} \\ 
    \hline
    vic & atk & test acc & atk acc & num & test acc & atk acc & num & test acc & atk acc \\
    \hline
    9 & 0 & 98.5 & 100 & 35 & 96.2 & 100 & 40 & 95.7 & 99.9 \\
    9 & 1 & 98.2 & 100 & 37 & 96.2 & \textbf{0.4} & 39 & 93.8 & \textbf{0.6} \\ 
    9 & 2 & 98.6 & 100 & 36 & 96.5 & \textbf{0.1} & 37 & 95.8 & \textbf{0.2} \\ 
    9 & 3 & 98.6 & 99.6 & 32 & 96.2 & \textbf{6.7} & 34 & 96.1 & \textbf{2.3} \\ 
    9 & 4 & 98.7 & 100 & 34 & 94.7 & 100 & 35 & 95.2 & \textbf{0.5} \\ 
    9 & 5 & 98.4 & 99.9 & 39 & 95.7 & 99.9 & 39 & 95.7 & 99.9 \\ 
    9 & 6 & 98.1 & 99.8 & 33 & 95.6 & 100 & 32 & 96.6 & 99.9 \\ 
    9 & 7 & 98.8 & 99.9 & 37 & 96.7 & 98 & 37 & 96.1 & 99.2 \\ 
    9 & 8 & 98.2 & 99.7 & 36 & 96.2 & 99.5 & 36 & 95.5 & 99.7 \\ 
    \hline
    0 & 9 & 98.4 & 100 & 40 & 95.3 & \textbf{4.1} & 39 & 96 & \textbf{3.4} \\ 
    1 & 9 & 98.7 & 99.7 & 38 & 95.9 & 99.9 & 38 & 96.4 & \textbf{0.3} \\ 
    2 & 9 & 98.4 & 99.8 & 32 & 97.5 & 100 & 30 & 97.2 & 100 \\ 
    3 & 9 & 97.2 & 99.8 & 30 & 90.6 & 100 & 32 & 95.8 & 99.3 \\ 
    4 & 9 & 97.9 & 100 & 37 & 94.1 & 100 & 38 & 93.6 & 100 \\ 
    5 & 9 & 98.6 & 100 & 34 & 96.6 & \textbf{0.8} & 34 & 96.7 & \textbf{0.7} \\ 
    6 & 9 & 98.6 & 99.9 & 36 & 97 & 99 & 33 & 96.8 & 100 \\ 
    7 & 9 & 98.8 & 99.7 & 36 & 96.9 & 97.6 & 35 & 97.2 & 98.4 \\ 
    8 & 9 & 98.3 & 100 & 34 & 96.3 & 97.8 & 37 & 95 & 99.5 \\ 
    \hline
\end{tabular}
\label{table:pruning_neurons_performance}
\vspace{-3ex}
\end{table*}

\begin{table}[ht]
\caption{Experiments with only Adjusting Extreme Weights}
\vspace{-2ex}
\centering
\begin{tabular}{|c|c|c|c|c|}
    \hline
    \multicolumn{2}{|l|}{Target} & \multicolumn{3}{|l|}{Adjusting Extreme Weights} \\ 
    \hline
    vic & atk & num & test acc (\%) & atk acc (\%) \\
    \hline
    9 & 0 & 30 & 98.2 & 0.4 \\
    9 & 1 & 34 & 98.2 & 2.8 \\ 
    9 & 2 & 37 & 97.5 & 0.1 \\ 
    9 & 3 & 24 & 98.2 & 0.3 \\ 
    9 & 4 & 28 & 97.1 & 2.8 \\ 
    9 & 5 & 36 & 98.6 & 8 \\ 
    9 & 6 & 29 & 98.4 & 0.1 \\ 
    9 & 7 & 38 & 98.5 & 1.9 \\ 
    9 & 8 & 29 & 98.3 & 0 \\ 
    \hline
    0 & 9 & 30 & 98.6 & 0.3 \\ 
    1 & 9 & 28 & 97.3 & 19.3 \\ 
    2 & 9 & 31 & 98.4 & 0 \\ 
    3 & 9 & 31 & 98.6 & 0 \\ 
    4 & 9 & 31 & 98.3 & 10.1 \\ 
    5 & 9 & 18 & 97.8 & 1.5 \\ 
    6 & 9 & 38 & 98.5 & 0.2 \\ 
    7 & 9 & 33 & 98.2 & 1.2 \\ 
    8 & 9 & 25 & 98.5 & 7.9 \\ 
    \hline
\end{tabular}
\label{table:pruning_extreme_values_performance}
\vspace{-2.5ex}
\end{table}

\subsection{Defense Method Evaluation}
Next we show the experiment results of our pruning method under different circumstances. We conduct experiments on the MNIST dataset to show that both the neuron pruning process and the process of adjusting the extreme weights are all essential parts in our defense method. Further experiments on the fine-tuning process are also reported here. 
Then, we test the whole defense procedure on the Fashion-MNIST dataset and present the promising results. 
In the end, we use the same defense procedure on the CIFAR-10 dataset under the state-of-the-art Distributed Backdoor Attack and prove the effectiveness of our method. 

\subsubsection{Pruning Neurons}
According to the definition of our threat model, the goal of the attacker is to manipulate the prediction results of some inputs $x_{i}'$ with backdoor patterns to another label. To be more specific, those inputs $x_{i}$ should be predicted as label $y_{i} = T$ without backdoor patterns. However, after adding the backdoor patterns to the original images, $x_{i}'$ will be predicted as label $y_{i}' = F$ by the compromised model. Under this attack, the backdoor patterns and the victim label $T$ and target label $F$ should all be determined by the attacker. 

The work of Liu et al. \cite{DBLP:conf/raid/0017DG18} has shown that pruning dormant neurons can effectively mitigate the backdoor attack success rate. However, when we pruned neurons in federated learning scenarios, our experiments showed that the success rate of mitigating backdoor attacks depends on the victim label $T$ and target label $F$ selected by the attacker. 
Detailed experiment results of neurons pruning process on distributed MNIST dataset when attackers have different attack targets are represented in Table \ref{table:pruning_neurons_performance}.
\textit{vic} stands for the victim label that the attackers want to attack. \textit{atk} stands for the target label that the attackers want the backdoor data being predicted. \textit{test acc} stands for the performance of model on test dataset. \textit{atk acc} stands for the performance of model on backdoor dataset. Backdoor dataset is composed of images that originally belong to \textit{vic} label in the test dataset been added backdoor patterns. \textit{num} stands for the number of neurons that are pruned during this process. There are 50 neurons in this convolutional layer.

With the Ranking Vote method, the backdoor attack success rates drop below $10\%$ only in 5 out of 18 cases shown in \ref{table:pruning_neurons_performance}. With the Majority Vote method, the chance of successful defense is $38.9\%$ (7 out of 18). 
These results show that pruning neurons only is not able to mitigate backdoor attacks effectively. 
Figure \ref{fig:neuron_pruning_methods_comp} shows that 
the attack success rate will not decrease until the test accuracy drops to an unacceptable level. 
This is because some neurons that support backdoor patterns may also perform an essential function in supporting benign inputs.

\begin{table*}[h]
\caption{Experiments of Fine-tuning Process on MNIST Dataset}
\vspace{-2ex}
\centering
\begin{tabular}{|c|c|c|c|c|c|c|c|c|c|}
    \hline
    \multicolumn{2}{|l|}{Attacker Target} & \multicolumn{2}{|l|}{Training Phase} & \multicolumn{2}{|l|}{Pruning without Fine-tuning} & \multicolumn{2}{|l|}{Pruning with Fine-tuning} \\ 
    \hline
    vic label & atk label & test acc (\%) & atk acc (\%) & test acc (\%) & atk acc (\%) & test acc (\%) & atk acc (\%) \\
    \hline
    9 & 0 & 98.2 & 99.7 & 94.7 & 0.3 & 96.6 & 0.8 \\
    9 & 1 & 98.6 & 100 & 95.1 & 0.4 & 96.9 & 0.1 \\
    9 & 2 & 98.2 & 99.9 & 95.7 & 4.1 & 97.2 & 2.5 \\
    9 & 3 & 98.3 & 99.6 & 95.6 & 36.9 & 97.2 & 8.8 \\
    9 & 4 & 98.5 & 100 & 97.2 & 5 & 98.1 & 3.5 \\
    9 & 5 & 98.5 & 99.9 & 86.6 & 14 & 94.6 & 2.8 \\
    9 & 6 & 98.6 & 99.7 & 96.4 & 0.1 & 97.5 & 0.1 \\
    9 & 7 & 98.9 & 99.8 & 94.1 & 0.8 & 98.2 & 0.5 \\
    9 & 8 & 98.5 & 99.9 & 93.9 & 4.5 & 96.6 & 3.1 \\
    \hline
    0 & 9 & 97.8 & 99.8 & 95.4 & 0 & 96.9 & 0.9 \\
    1 & 9 & 98.2 & 99.5 & 94.5 & 0.1 & 97.2 & 0.5 \\
    2 & 9 & 98.3 & 100 & 94.9 & 1.2 & 97.5 & 0.5 \\
    3 & 9 & 98.5 & 99.5 & 95 & 34.9 & 96 & 26.3 \\
    4 & 9 & 98.5 & 99.9 & 93.4 & 27.1 & 96.4 & 14.7 \\
    5 & 9 & 98.5 & 99.8 & 95.2 & 1.3 & 97.5 & 1.6 \\
    
    6 & 9 & 98.3 & 99.3 & 94.8 & 1.5 & 97.3 & 0.3 \\
    7 & 9 & 97.9 & 99.1 & 93.8 & 18.5 & 96.9 & 16.4 \\
    8 & 9 & 97.5 & 99.4 & 93.3 & 1.2 & 96.3 & 0.5 \\
    \hline
\end{tabular}
\label{table:pruning_fine_tuning}
\vspace{-1ex}
\end{table*}

\begin{table*}[h]
\caption{Experiments on Fashion-MNIST Dataset}
\vspace{-2ex}
\centering
\begin{tabular}{|c|c|c|c|c|c|c|c|c|c|c|c|}
    \hline
    \multicolumn{2}{|l|}{Target} & \multicolumn{2}{|l|}{Training Phase} & \multicolumn{2}{|l|}{Pruning Neurons} & \multicolumn{2}{|l|}{Adjusting Extreme Weights} & \multicolumn{2}{|l|}{Defense with Fine-tuning} \\ 
    \hline
    vic & atk & test acc & atk acc & test acc & atk acc & test acc & atk acc & test acc & atk acc \\
    \hline
    9 & 0 & 88.8 & 99.8 & 83.2 & 2.8 & 82.9 & 2.1 & 86.3 & 9.0 \\
    9 & 1 & 88.7 & 99.4 & 82.6 & 6.5 & 82.1 & 0 & 87.0 & 0 \\
    9 & 2 & 87.8 & 99.8 & 82.2 & 2.7 & 82.1 & 3.1 & 85.6 & 12.6 \\
    9 & 3 & 87.5 & 99.6 & 84.4 & 0.3 & 84.4 & 0 & 86.2 & 0.4 \\
    9 & 4 & 87.8 & 99.7 & 81.0 & 2.9 & 80.6 & 0 & 85.8 & 2.3 \\
    9 & 5 & 88.6 & 99.7 & 81.2 & 11.9 & 80.6 & 3.6 & 86.1 & 10.4 \\
    9 & 6 & 86.3 & 99.8 & 82.8 & 93.6 & 82.0 & 0.2 & 86.0 & 3.1 \\
    9 & 7 & 88.5 & 99.7 & 82.9 & 4.3 & 83.1 & 4.6 & 87.0 & 17.1 \\
    9 & 8 & 88.8 & 99.9 & 84.7 & 87.7 & 85.0 & 3.2 & 87.2 & 2.9 \\
    \hline
\end{tabular}
\label{table:fmnist_results}
\end{table*}

\begin{figure*}
    \centering
    \includegraphics[width=1\textwidth]{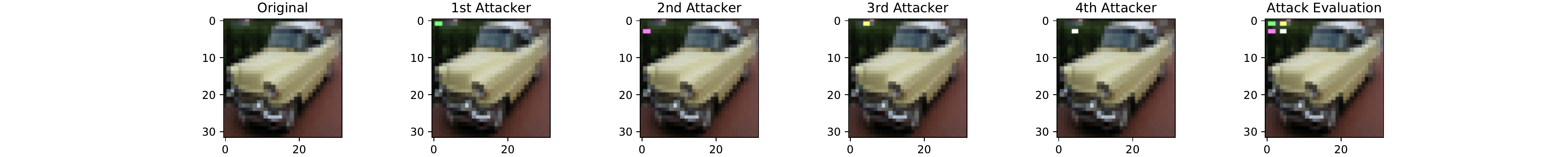}
    \vspace{-4ex}
    \caption{\scriptsize An original image from the CIFAR-10 dataset, followed by four different attackers each provides a portion of the attack pattern and the final attack evaluation will use the full attack pattern in the last image.
    }
    \label{fig:cifar10_attack_patterns_graph}
    \vspace{-2ex}
\end{figure*}

\subsubsection{Adjusting Extreme Weights}
Above we showed that pruning dormant neurons alone in federated learning is not enough to mitigate backdoor attacks. Next we show that adjusting extreme weights alone can mitigate backdoor attacks when the model structure is concise enough. Specifically, we are using a two-layer convolutional neural network with the first layer containing 8 neurons and the second layer containing 16 neurons to train on MNIST. 
Table \ref{table:pruning_extreme_values_performance} represents
experiment results of adjusting extreme weights process on distributed MNIST dataset when attackers have different attack targets. 
The success rate of backdoor attacks decrease from over $99\%$ to an average of $3.2\%$ while the accuracy of the model on test datasets remains almost the same. 
However, if we perform this process on a larger network (e.g.: two-layer convolutional network with the first layer containing 20 neurons, and the second layer containing 50 neurons), the attack success rate will not decrease. The reason could be that the backdoor training samples are leveraging lots of redundant neurons to reverse the correct prediction results while those backdoor neurons do not necessarily have extreme weights since they can dominate through numbers.
So, pruning neurons is necessary when there are redundant neurons in the network that can be leveraged by the backdoor attacks.

\subsubsection{Fine-tuning}
Since the neuron pruning process could not eliminate all the backdoor neurons, we can further improve the model performance on a normal test dataset by fine-tuning the pruned models through federated learning. We apply the same procedure in this step as we did in the previous training process. The only difference is that the model structure has been changed due to the neuron pruning process. During the federated fine-tuning, the attack success rate will also increase along with the improvement of test accuracy on the testing dataset, since attackers also participate in this process. However, after we perform the process of adjusting extreme weights in the end, we can find that the attack success rate will drop again to the same level as without the fine-tuning process, while the test performance will be much higher. 
Experiments on the MNIST dataset have shown that without the fine-tuning process, the average test accuracy on the validation dataset drops from $98.3\%$ to $94/4\%$, and the average attack success rate drops from $99.7\%$ to $8.4\%$. With the fine-tuning process, the average test accuracy on the validation dataset only drops to $96.9\%$, and the average attack success rate drops to $4.7\%$.
Detailed experiment results of the comparison between with and without fine-tuning process are reported in Table \ref{table:pruning_fine_tuning}. \textit{vic label} stands for the victim label that the attackers want to attack. \textit{atk label} stands for the target label that the attackers want the backdoor data being predicted. \textit{test acc} stands for the performance of model on test dataset. \textit{atk acc} stands for the performance of model on backdoor dataset. Backdoor dataset is composed of images that originally belong to \textit{vic label} in the test dataset been added backdoor patterns.

\begin{table*}[h]
\caption{Experiments on CIFAR-10 Dataset under Distributed Backdoor Attack}
\vspace{-2ex}
\centering
\begin{tabular}{|c|c|c|c|c|c|c|c|c|c|c|c|}
    \hline
    \multicolumn{2}{|l|}{target} & \multicolumn{2}{|l|}{training phase} & \multicolumn{2}{|l|}{pruning neurons} & \multicolumn{2}{|l|}{adjusting extreme weights} & \multicolumn{2}{|l|}{defense with fine-tuning} \\ 
    \hline
    vic & atk & test acc & atk acc & test acc & atk acc & test acc & atk acc & test acc & atk acc \\
    \hline
    9 & 0 & 73.0 & 88.5 & 72.7 & 59.9 & 71.6 & 21.2 & 71.8 & 39.6 \\
    9 & 1 & 71.6 & 87.0 & 71.0 & 87.6 & 70.9 & 18.1 & 71.1 & 19.2 \\
    9 & 2 & 72.9 & 84.0 & 71.7 & 73.3 & 70.4 & 26.6 & 71.6 & 29.6 \\
    9 & 3 & 72.7 & 87.7 & 71.4 & 12.0 & 71.3 & 13.1 & 71.0 & 45.8 \\
    9 & 4 & 72.9 & 85.3 & 71.9 & 3.3 & 70.5 & 4.1 & 71.5 & 49.1 \\
    9 & 5 & 72.8 & 85.5 & 72.4 & 4.4 & 71.8 & 4.8 & 71.5 & 46.9 \\
    9 & 6 & 71.0 & 91.3 & 71.7 & 82.6 & 70.5 & 7.9 & 72.2 & 5.8 \\
    9 & 7 & 72.1 & 91.2 & 72.2 & 85.7 & 71.3 & 8.5 & 71.5 & 8.8 \\
    9 & 8 & 72.6 & 88.2 & 72.4 & 10.8 & 71.2 & 12.8 & 71.5 & 49.9 \\
    \hline
\end{tabular}
\label{table:cifar10_results}
\vspace{-3ex}
\end{table*}

\subsubsection{Experiments on Fashion-MNIST}
Experiment results of the training process, neuron pruning process, extreme weights adjusting process and pruning with fine-tuning process on distributed Fashion-MNIST dataset are shown in Table \ref{table:fmnist_results}. 
The backdoor pattern used in this experiment is single-pixel backdoor pattern.
We have a total of 10 clients and use a 3-label data distribution among all the clients. There is one attacker among the clients. The global model has three convolutional layers and another two fully connected layers. We tested our defense method under different attacking targets selected by the attacker. According to the result, we can also notice that the neuron pruning process can only mitigate backdoor attacks targeting certain labels, but the following adjusting extreme weights can mitigate all backdoor attacks targeting any labels. On the other hand, the fine-tuning process can significantly improve the model performance on the main task, although it may also benefit the attacker since the attacker is also among the group of clients participating in the fine-tuning. 
Without a fine-tuning process, the average attack success rate drops from $99.7\%$ to $1.9\%$, and the test accuracy on the validation dataset drops from $88.1\%$ to $82.5\%$. With the fine-tuning process, the test accuracy on the validation dataset rises up to $86.4\%$, although the average attack success rate also increases from 1.9\% to $6.4\%$, as a tradeoff between performance and security.

\subsubsection{Experiments on CIFAR-10}
On CIFAR-10 dataset, we use the same attacking strategies as provided in the \textit{Distributed Backdoor Attack} \cite{DBLP:conf/iclr/XieHCL20}. The attack decomposes a global backdoor pattern into separate local patterns and embed them into the training set of different attackers respectively, as shown in Figure \ref{fig:cifar10_attack_patterns_graph}.
Experiment results are shown in Table \ref{table:cifar10_results}. 
We have a total of 10 clients and use a 3-label data distribution among all the clients. There are four attackers among the clients.
Despite of different attacking strategies, the experiment results are very similar to what we have observed in the experiments on Fashion-MNIST. 
The neuron pruning process can mitigate some but not all backdoor attacks targeting different labels. 
The following adjusting extreme weights can further mitigate all backdoor attacks targeting different labels.
In average, the attack success rate drops from $87.6\%$ to $13.0\%$ and the test accuracy on the validation dataset drops from $72.4\%$ to $71.0\%$. With fine-tuning process, the average attack success rate drops $54.9\%$ while the average test accuracy only drops $0.9\%$.

\begin{figure}
    \vspace{-2.5ex}
    \centering
    \includegraphics[width=0.45\textwidth]{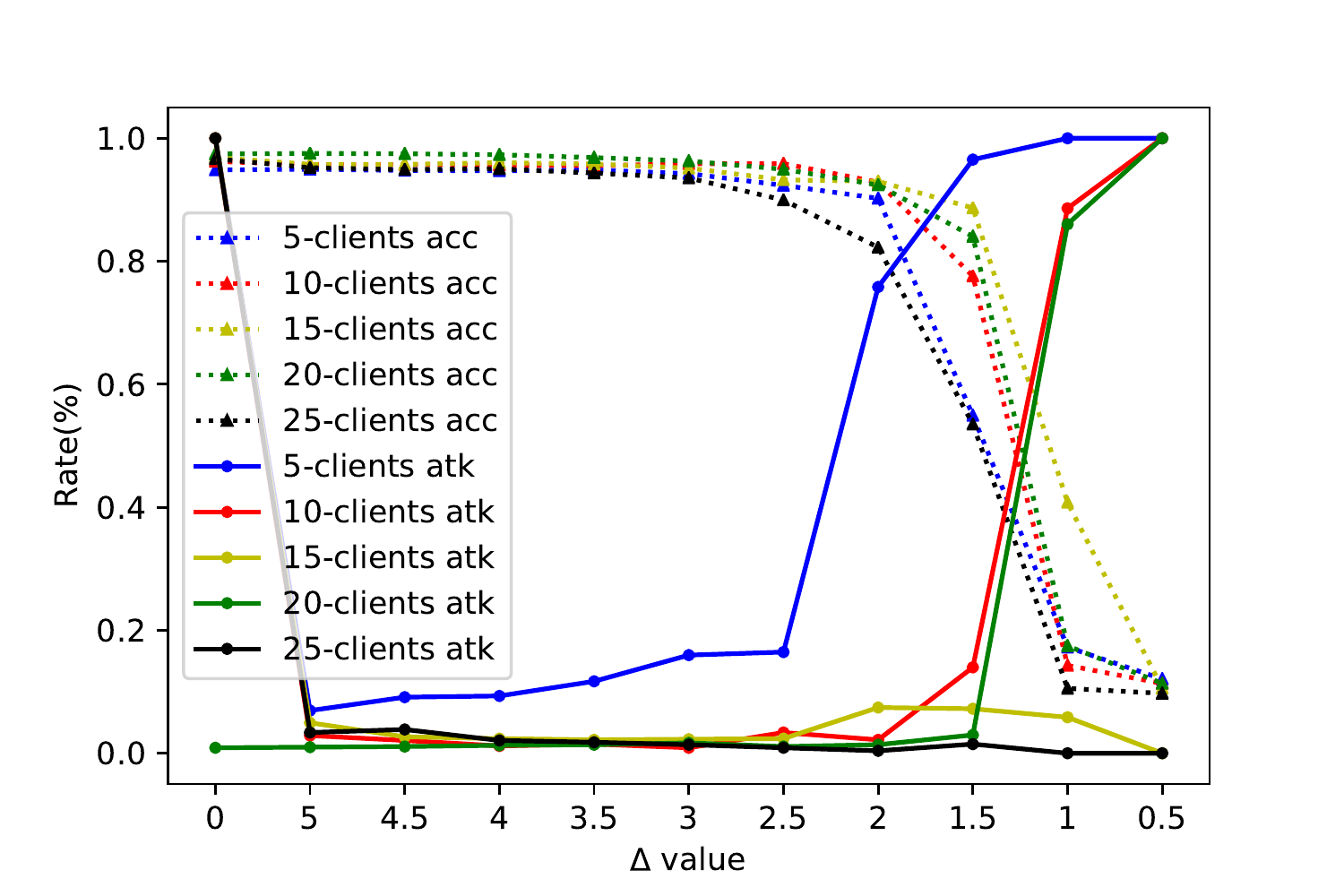}
    \vspace{-2ex}
    \caption{The test accuracy of model on validation dataset and backdoor attack success rate of model during the adjusting extreme values phase with different number of clients selected in each round in the training process. 
    The model is trained on MNIST dataset in 3-label distribution among 50 clients.}
    \label{fig:random_select_adjusting}
    \vspace{-2ex}
\end{figure}

\subsection{Randomly Selected Clients Evaluation}

Previous experiments are all tested with 10 clients and each client participate in every round of training process. This is a simplified process compared with original federated learning definition \cite{DBLP:conf/aistats/McMahanMRHA17} as we discussed before. 
In this section, we perform experiments on a more realistic federated learning scenario with 50 clients and $10\%$ of them ($5$ clients) are attackers. We randomly select different number of clients (5, 10, 15, 20 and 25 clients, respectively) in each experiment and observe the performance of our defense method. We use the MNIST dataset under 3-label distribution for this comparison experiments. 
From Fig \ref{fig:random_select_adjusting}, we can see that the performance of the model behave very similar to each other, although they are trained with randomly selected different number of clients. 
With reasonable number of pruned neurons and $\Delta$ values (same settings as previous experiments), the attack success rate can still be greatly mitigated using our defense method.

\section{Discussion}

\begin{figure}
    \vspace{-2.5ex}
    \centering
    \includegraphics[width=0.45\textwidth]{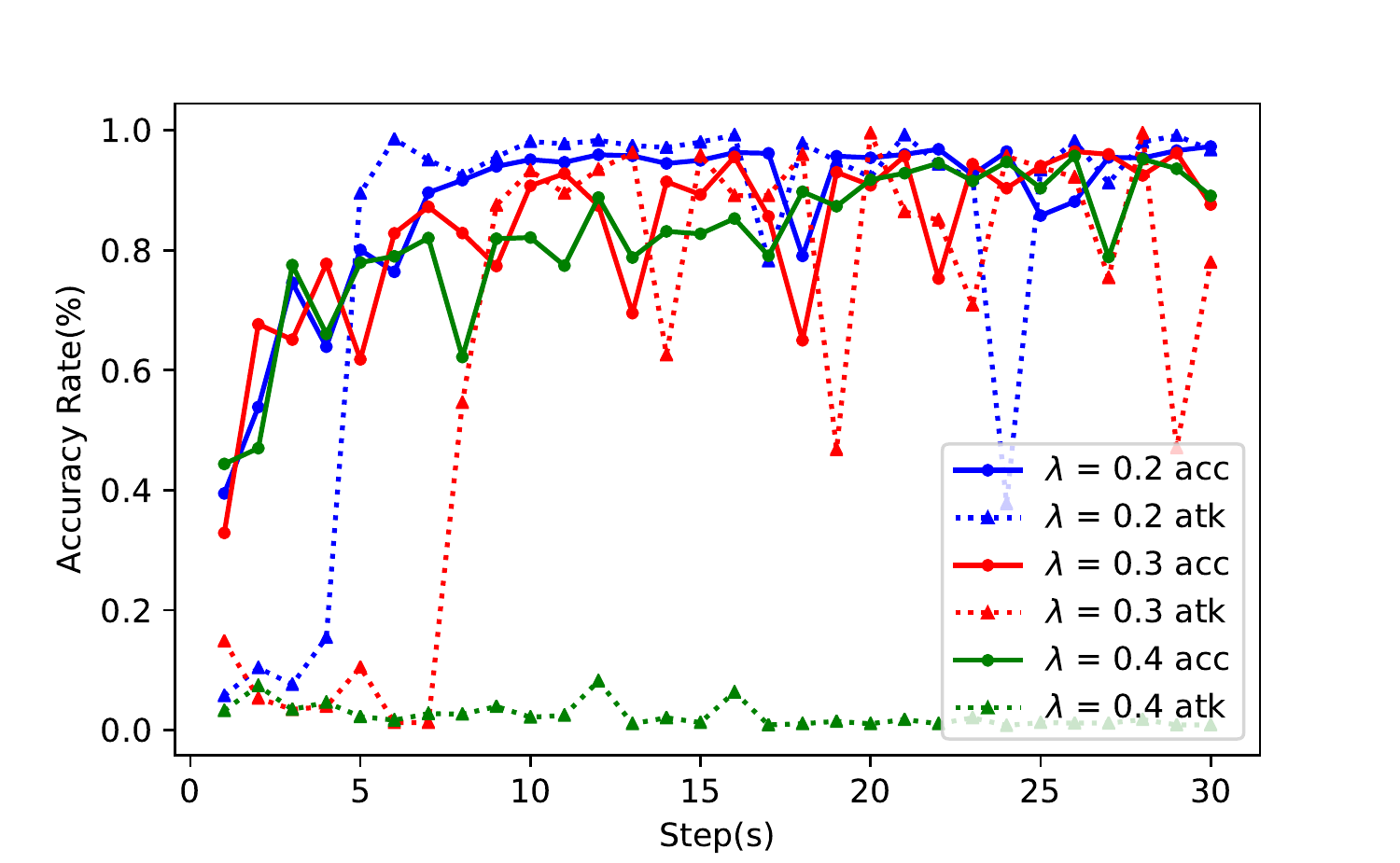}
    \vspace{-2ex}
    \caption{The training process of model with different regularization coefficient $\lambda$ in the last convolutional layer. The model is trained on MNIST dataset in 3-label distribution among 10 clients. The training processes are identical with the same data distribution, 30 epochs of training time and the same attacker's amplification update coefficient $\alpha = 3$. Solid line stands for the accuracy rate and dotted line stands for the attack success rate.}
    \label{fig:model_regularization}
    \vspace{-2ex}
\end{figure}

\subsection{Regularization Method}
\label{sec:regularization_method}
The extreme weights adjusting method tries to eliminate those extreme values in the network weights. This method is very similar to the idea of regularization. However, we find that directly applying the L2 regularization penalty of all layers on the loss function will potentially degrade the model's performance on the test dataset. Instead, using the L2 regularization penalty only on the last convolutional layer can increase the network's robustness against a backdoor attack.
Related experiment results can be seen on Figure \ref{fig:model_regularization}.
After adding the regularization method in the last convolutional layer, it becomes much more challenging for the attackers to compromise the global model with a sufficient large regularization factor. 
However, such robustness of the model comes with a loss of performance on the designed tasks (test accuracy on the testing dataset). So, there is a trade-off between the robustness against backdoor attacks and the performance of the model. A larger regularization coefficient provides more robustness but will suffer from a loss in the test performance.

\subsection{Model Architectures}
The neuron pruning method tries to simplify the model architectures and ensure that every remaining neuron is essential to the designed task (measured by the test accuracy on the testing dataset).
Experiments (in Table \ref{table:pruning_extreme_values_performance}) have shown that when the model is concise/simple enough, we can even skip the neuron pruning step, while simply adjusting extreme weights in the last convolutional layer can mitigate backdoor attacks. 
However, it is almost impossible to design a network architecture that perfectly matches the least requirement of the complexity with an unseen dataset. Simple architectures may cause the model to suffer from large bias errors, while complex architectures will make the model more vulnerable to backdoor attacks. 
That is why we introduced the neuron pruning process to leverage the information from all clients to prune as many unnecessary neurons as possible without hurting the model's performance on its designed tasks. 

\subsection{Possible Attacks}
There is and will always be an arms race between attackers and defenders. In this section, we will briefly discuss some possible attacks on our defense mechanism. 
Firstly, we consider two different attacks on the federated pruning process. 
Recall that the assumption of the backdoor neurons is that the backdoor attacks will trigger those neurons that are typically not being activated by the normal input.
Naturally, we can form two kinds of attacks based on this assumption. The first one will try to attack the neurons ranking algorithm, such that the backdoor neurons ranked higher than those essential neurons used by normal input. So, the backdoor neurons will not be removed until the test accuracy on the validation dataset drops below a certain threshold. 
The second attack method will try not to generate such backdoor neurons and use the essential neurons being used by the other normal input to trigger the backdoor pattern, which is called ``pruning-aware attack'' \cite{DBLP:conf/raid/0017DG18}. 
To achieve this, we assume the attacker could obtain the final pruning mask created by all the clients (which is nearly impossible in real cases). The attacker would then use this pruning mask in the local training process to avoid using the pruned neurons. Moreover, the attacker can enforce the backdoor neurons are the same as the essential neurons used by other normal inputs.
We empirically studied the influence of these two attacks on our federated pruning algorithm. Such attacks nearly do not influence the defense results (attack success rate will still significantly decrease). The reason could be that the attackers still take the minority group ($10\%$) in our federated system settings. 
Besides, by using ranking numbers or voting masks in our pruning algorithm, the influence of some updates of manipulated rankings is trivial. What is more, the pruning extreme values process will further mitigate the influence of ``pruning-aware attack'' as we have shown in the previous experiments.

Secondly, we consider attacks to pruning extreme values. If we assume the attackers are aware of the pruning extreme values process, they could self-prune extreme values during the training process. In this way, when the server prunes extreme values, it will not affect the backdoor attack success rate since the backdoor attacks will not rely on extreme values. 
However, in our experiments, we find that the federated pruning process will significantly mitigate such attacks. 
The reason could be that by limiting extreme values during the training process, the backdoor tasks will have to leverage the dormant neurons to launch backdoor attacks. At the same time, our federated pruning process will restrain the use of dormant neurons. Thus, our defense method is robust under these attacks.

\section{Conclusion}
We proposed a new method to mitigate backdoor attacks in federated learning. 
We can simplify the model architectures through the federated neuron pruning process while maintaining good performance of the model on designed tasks. Then, adjusting extreme weights in the simplified model can effectively degrade the success rate of backdoor tasks. We evaluated our method with different attack settings (attack targets/labels, backdoor patterns, data distribution among clients, and datasets). Our experiments also showed that the federated fine-tuning process after pruning neurons could further improve the model performance on designed tasks without hurting the defending method's performance after adjusting the extreme values process. 

\bibliographystyle{abbrv}
\bibliography{main}

\end{document}